# High energy terahertz pulses from organic crystals: DAST and DSTMS pumped at Ti:sapphire wavelength

B. Monoszlai[1,*], C. Vicario[1], M. Jazbinsek[2] and C.P. Hauri[1,3,*]

[1]*Paul Scherrer Institute, 5232 Villigen PSI, Switzerland*
[2]*Rainbow Photonics AG, 8048 Zürich, Switzerland*
[3]*Ecole Polytechnique Federale de Lausanne, 1015 Lausanne, Switzerland*

Abstract
High energy terahertz pulses are produced by optical rectification (OR) in organic crystals DAST and DSTMS by a Ti:sapphire amplifier system centered at 0.8 µm. The simple scheme provides broadband spectra between 1 and 5 THz, when pumped by collimated 60 fs near-infrared pump pulse and it is scalable in energy. Fluence-dependent conversion efficiency and damage threshold are reported as well as optimized OR at visible wavelength. © 2013 Optical Society of America

One of the recognized concepts for high-field Terahertz generation is based on efficient optical rectification (OR) in organic crystals of mid-infrared, femtosecond laser pulses [0-0]. While this THz conversion scheme routinely provides THz pulses of MV/cm field strength, it requires a sophisticated multi-mJ femtosecond mid-IR laser system. The shortage of efficient lasing materials in the wavelength range between 1.3-1.5 µm requires the OR pump to be produced by whitelight-seeded frequency mixing processes and optical parametric amplifiers, which themselves are pumped by an amplified Ti:sapphire laser [0]. For many applications, however, a less sophisticated laser system is favourable as pump. In fact, direct pumping of organic crystals by Ti:sapphire laser would significantly reduce the overall complexity of the THz production and opens the possibilities to a broader scientific audience. To our best knowledge, OR in organic crystals pumped by a multi-mJ Ti:sapphire system has not been considered for high-energy Terahertz pulse generation so far.

In this Letter we present results on THz generation at sub-µJ pulse energy by optical rectification of an amplified Ti:sapphire pulse in DAST and DSTMS. We performed systematic studies on fluence-dependent conversion efficiencies as well as spectral THz shapes, and give indications on the maximum field and damage threshold of the organic rectifier crystals for a 0.8 µm pump wavelength. We also present THz measurements close to the optimum visible wavelength for pumping organic crystal.

The experimental setup is illustrated in Fig. 1. For our studies we used a 10 mJ, 60 fs Ti:sapphire amplifier system at 100 Hz repetition rate. The beam is collimated prior to pumping the organic crystals at hand, being a 0.86 mm thick DAST and a 0.2 mm thick DSTMS with 4.5 and 3 mm free aperture, respectively. Unlike other rectification schemes no pulse front tilting [0] or optical parametric amplifiers are used in our experiment.

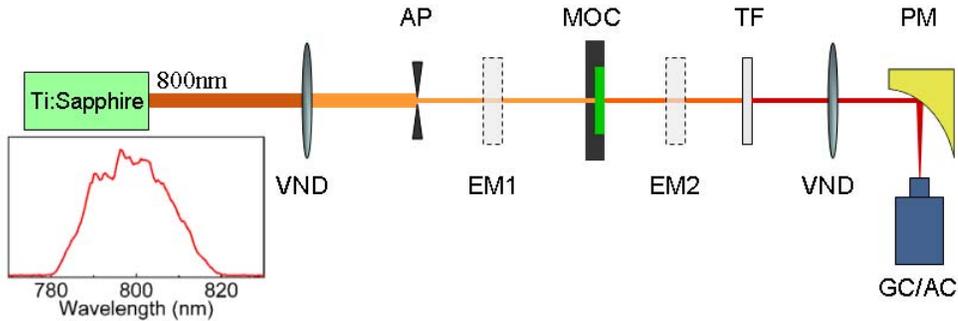

Fig. 1. Schematic of the THz generation setup for organic crystals pumped at 800 nm with the Ti:sapphire amplifier (100 Hz, 10 mJ). Variable ND filter wheel (VND), aperture (AP), mounted organic crystal (MOC), teflon filter (TF), parabolic mirror (PM), laser energy meter at two positions (EM#), and the diagnostics: Golay cell (GC) and autocorrelator (AC).

The collimated THz radiation is characterized in energy by a calibrated Golay cell, and in the time-frequency domain, by a first-order autocorrelation (AC). The latter offers a broadband bandwidth acceptance in the range of 0.1-20 THz to avoid spectral limitations associated to the crystal-based electro-optical sampling scheme. The instrument is based on a conventional Michelson-interferometer, with an additional parabolic mirror before the detector, what was also a Golay-cell in our case. Throughout the experiment a 2 mm thick Teflon filter is used to reduce the background of the pump laser to the detection level of the sensor. All the measurements have been performed at room temperature.

The crystal orientation is chosen such that the element $\chi^{OR}_{122}$ of the nonlinear susceptibility tensor is employed for optical rectification. While $\chi^{OR}_{111}$ offers most efficient THz generation when pumped with short wavelength infrared (SWIR) radiation between 1.3-1.5 μm (490 pV/m), the index $\chi^{OR}_{122}$ is best suited for near infrared pumping (700-800 nm), as predicted in [0] and experimentally confirmed in [0].

The highest conversion efficiency is provided for perfect velocity matching between the THz phase velocity and the group velocity of the optical pump pulse which is equivalent to being parallel to the crystal b-axis. For $\chi^{OR}_{122}$ at 800 nm a value of 166 pV/m is reported [0-0]. This indicates less efficient optical rectification than with a SWIR pump.

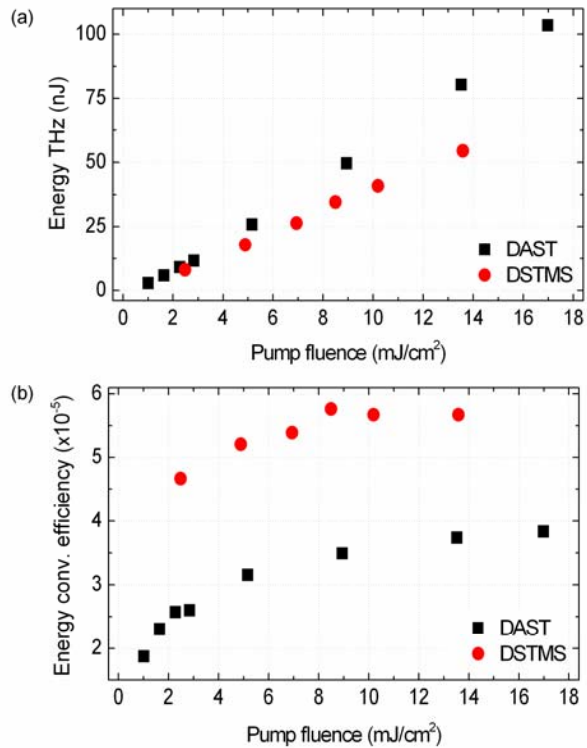

Fig. 2. Background-corrected Terahertz (a) pulse energy and corresponding (b) conversion efficiencies as function of the applied Ti:sapphire pump fluence for DAST and DSTMS, with emitting surface of 15.9 mm² and 7.1 mm² respectively. Measured after 2mm teflon plate.

Fig. 2 shows the THz pulse energy and the near IR-to-THz conversion efficiency as function of the pump fluence for DAST and DSTMS. The measured THz energy increases almost linearly with respect to the pump fluence to 100 nJ for DAST (15.9mm² emitter surface) and 50 nJ for DSTMS (7.1 mm² emitter surface) at 17 mJ/cm² and 14 mJ/cm² pump fluences respectively. Please note that these values were measured after 2mm of teflon filter, and are not corrected with the ~25% average absorption across the 0.3-5 THz range. Up to this fluence, which is close to the damage threshold, no saturation could be observed in the THz energy. The conversion efficiency, shown in Fig 2(b), slightly increases for higher pump fluence and levels of at 6e-5 for DAST and 4e-5 for DSTMS, respectively. Shown in Fig. 3 are typical THz spectra recorded for OR with the 800 nm pump. The output covers 1-5 THz and peaks around 2 THz.

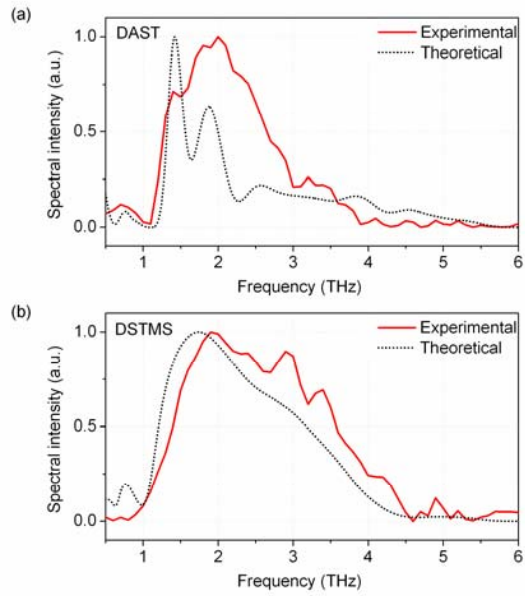

Fig. 3. THz spectra for (a) DAST and (b) DSTMS pumped at 800 nm (fluence 6 mJ/cm$^2$) along the b-axis. The theoretical results were corrected for the transmission of the 2-mm thick teflon sheet.

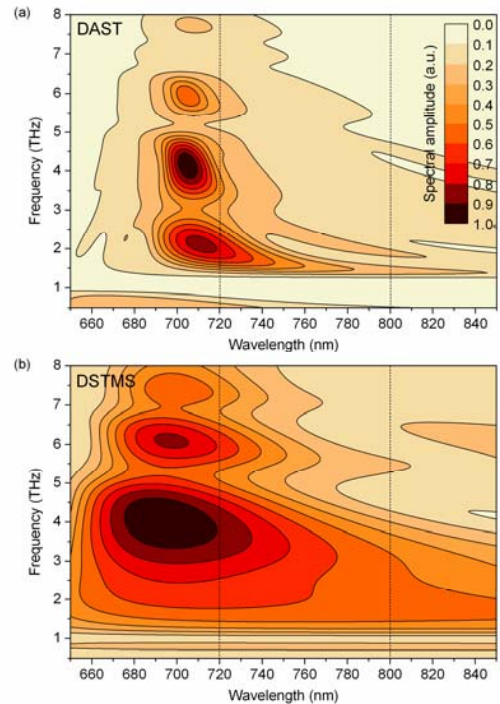

Fig. 4. Calculated spectral amplitudes of the THz frequency components (normalized), as a function of the pump wavelength, in the case of (a) 0.86 mm thick DAST and (b) 0.2 mm thick DSTMS. The vertical dashed lines show the pump laser wavelengths used in the presented investigation.

In the experiment, only a weak spectral content is observed at frequencies lower than the phonon-active mode at 1.1 THz. The overall spectral response is in acceptable agreement with the theoretically expected spectral output (dashed lines) which considers the crystal thickness, pulse duration and pump wavelength. Shown in Fig. 4 is an overview on the theoretically expected spectral content produced in DAST and DSTMS as function of the pump laser wavelength. The calculated curves are based on velocity matching and linear THz absorption in the crystal [0, 0]. We associate the small discrepancy between experimental and theoretical results, observed in particular for DAST (see Fig. 3a), to the fact that effects like nonlinear absorption and cascading are not treated in the numerical calculation.

The highest pump fluence used in the experiment is close to the measured damage threshold. For 68 fs pulses (FWHM) at 800 nm a damage threshold of 300 GW/cm$^2$ (20 mJ/cm$^2$) has been measured for DAST, and slightly lower values for DSTMS.

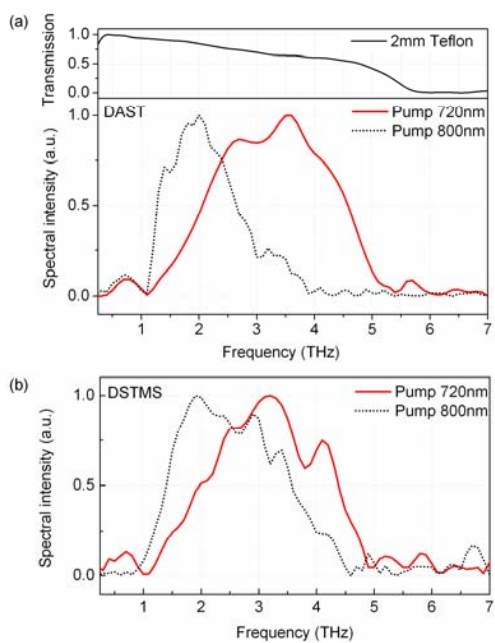

Fig. 5. Measured THz spectra for (a) DAST and (b) DSTMS pumped at 720 nm and 800nm. For visible pumping the spectrum is limited by the teflon transmission, cutting off for frequencies >5 THz, as shown in (a).

Although the electric field shape was not measured, the maximum field strength can be estimated by the pulse energy and its spectral content. With a strongly focused spot size an electric field of 1MV/cm [0] is calculated. This field strength has been shown sufficiently high to initiate nonlinear THz phenomena [0-0]. Previous works show that the focusability of THz radiation from organic crystal emitters is excellent, mainly due to the simple generation scheme. The resulting THz beam properties, such as a non-distorted pulse wavefront and identical divergence for both lateral dimensions help to achieve diffraction limited spot size. In principle, higher THz pulse energies become feasible when scaling the size of the Finally we present studies on visible wavelengths which are best suited to achieve broadband and efficient THz radiation.

In fact, as indicated by Fig. 4, the most effective visible pump laser wavelengths for broadband OR in organic crystals are located around 680-740 nm for DAST and around 700nm for DSTMS. Unfortunately these wavelengths are outside the typical gain bandwidth of a commercially available Ti:sapphire amplifier.

However, in order to verify the theoretical expectations, THz spectra were recorded for a pump wavelength of 720 nm for DAST and DSTMS, using the frequency-doubled signal from an OPA. As predicted by Fig. 4, under these conditions velocity matching is fulfilled also for higher THz frequencies resulting in multi-octave spanning spectra covering 0.5-5 THz. The corresponding experimental results are shown in Fig. 5. At this visible pump wavelength the expected conversion efficiency is 10 (2.5) times higher for DAST (DSTMS) compared to pumping at 800nm.

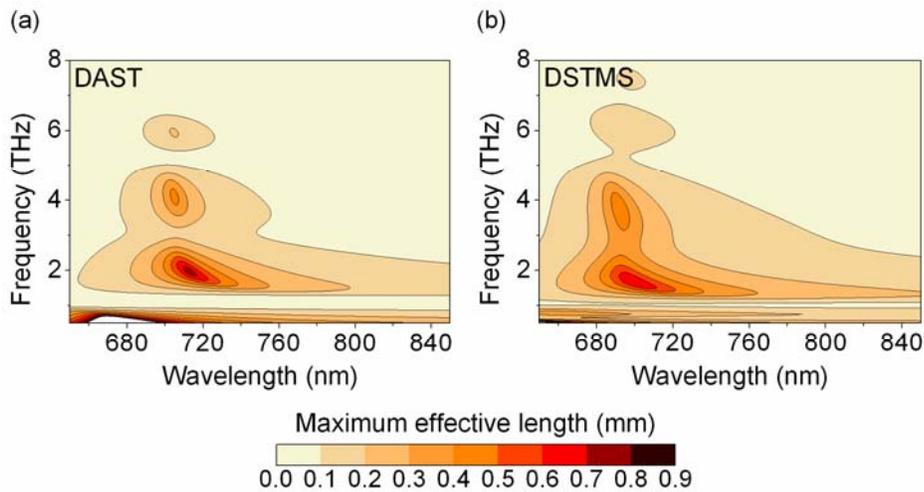

Fig. 6. Maximum effective length for efficient THz generation in (a) DAST and (b) DSTMS along the b-axis, for visible/nIR pumping wavelength.

Finally we would like to mention that the crystal thickness also has a strong influence on the rectified spectral components due to the phase-matching conditions. Fig. 6 gives an overview on the maximum effective lengths for THz-wave generation as defined in [0], as a function of the pump wavelength. With In conclusion, we have investigated the generation of high-energy THz pulses in organic crystals pumped by conventional Ti:sapphire laser. Broadband THz radiation up to 5 THz has been demonstrated with pulse energies up to 0.1 µJ, and damage thresholds were reported. Furthermore we demonstrated the possibility to generate higher THz frequencies by using a pump laser wavelength of 720nm. The measured efficiencies are comparable to other ordinary OR crystals, like the semiconductor ZnTe (3.1e-5) [0]. Even higher efficiencies could be expected for optimized crystal thickness. Although LiNbO$_3$ based sources show slightly better conversion efficiencies [0], the organic crystal allows to access significantly higher frequencies, well beyond 1 THz, and work without pulse front tilting. Due to its simplicity the presented source is of interest for both spectroscopic and high-field applications. It furthermore opens new opportunities in scaling THz sources towards higher field strength, thanks to the increasing availability of commercial Ti:sapphire systems at the TW power level.

We are grateful to C. Medrano and B. Ruiz from Rainbow Photonics for fruitful discussions. This work was supported by NFS (grant no 51NF40-144615) in the framework of NCCR-MUST and SwissFEL. B. M. acknowledges support from the Sciex-NMS, (Grant no. 12.159). CPH acknowledges support from the Swiss National Science Foundation (grant no. PP00P2_128493).